\RequirePackage{lineno}
\documentclass[%
 reprint,
twocolumn,
showpacs,preprintnumbers,showkeys,
 amsmath,amssymb,
 aps,
prc,
floatfix,
]{revtex4-2}

\usepackage{graphicx}% Include figure files
\usepackage{dcolumn}% Align table columns on decimal point
\usepackage{bm}% bold math
\usepackage{amssymb}
\usepackage{float}
\usepackage[caption = false]{subfig}
\usepackage{hyperref}% add hypertext capabiliti
\begin{document}

%\preprint{APS/123-QED}

\title{Investigating the role of partonic and hadronic dynamics in mass splitting of elliptic anisotropy in p-Pb collisions at $\sqrt{s_{NN}}$ = 5.02 TeV}
\author{Debojit Sarkar}
\email{debojit03564@gmail.com}
\author{Subikash Choudhury}
\author{Subhasis Chattopadhyay}
%\email{sub@vecc.gov.in}
\affiliation{Variable Energy Cyclotron Centre, HBNI, 1/AF-Bidhannagar, Kolkata-700064, India}
%\date{\today}

%\corauth[cor]{Corresponding author.}
%\ead{sub.chattopadhyay@gmail.com}
%\address[label1]{Variable Energy Cyclotron Centre, 1/AF-Bidhannagar, Kolkata-700064, India}

\begin{abstract}
The mass ordering of $v_{2}^{hadron}$ is regarded as one of the key signatures of collective behaviour in ultra relativistic heavy ion collisions. This observation has been found to be in compliance with the hydrodynamical response of a strongly interacting system to the initial spatial anisotropy. Flow co-efficients measured with identified particles in p-Pb/d-Au collisions have  shown similar mass-splitting of $v_{2}^{hadron}$ indicating towards the presence of collective dynamics in small collision systems. Arguably, small size in the overlap geometry of such colliding systems may not be suitable for hydrodynamical treatment that demands
an early thermalization. Studies based on a multi phase transport model suggests that elliptic or triangular anisotropy
is primarily due to escape mechanism of partons rather than hydro like collectivity and mass ordering of $v_{2}^{hadron}$ can be generated from coalescence dynamics as implemented in string melting version of AMPT even when parton azimuthal directions are randomized. In this work, studies have been performed on p-Pb collisions at $\sqrt{s_{NN}}$ = 5.02 TeV using AMPT model which has been found to explain the elliptic and traingular flow in such a system where escape mechanism is the dominant source of flow generation. We report that the mass splitting of $v_{2}^{hadron}$ can originate independently both at the partonic and hadronic level in the string melting version of the AMPT model.

%We report that a modification in the initial phase-space distribution of the partons  would be required prior %to hadronization by quark coalescence to generate mass ordering of $v_{2}^{hadron}$ in AMPT.

\end{abstract}

%\pacs{24.30.Cz, 29.40.Mc, 24.60.Dr.}
%\begin{keyword}


%\end{keyword}
%\end{frontmatter}

\maketitle

\section{Introduction}
Wealth of data collected during the decades of operation at the RHIC and first few years at the LHC provide compelling evidence that a strongly interacting and nearly
perfect fluid of quarks and gluons are produced in Pb-Pb/Au-Au collisions at ultra-relativistic energies. A manifestly evident signature for the formation of such a matter is  the collective motion (flow) of the final state particles.  A  large azimuthal anisotropy in the momentum space
has been regarded as one of the most definitive and strong indication of such collective behaviour and argued to be a consequence of collective
expansion of the system that starts with an initial azimuthal anisotropy in the coordinate space. 
%The over-lap area in a non-central A-A collisions has a spatial anisotropy. 
%It has been identified that final state momentum anisotropy is a hydrodynamic response to the initial %spatial an-isotropy in the overlap geometry of the colliding nuclei. 
This interpretation was initially applicable to heavy-ion collisions as large system size and high density were considered to be mandatory for the creation of a thermalized deconfined medium.
Elliptic anisotropy, long-range ridge structures, mass ordering of $v_{2}$, baryon to meson enhancement at intermediate $p_{T}$ etc which were once attributed to the hydro-dynamical evolution of a strongly interacting system of large dimensions found to be challenged when analogous measurements in small collision systems produced similar outcome \cite {alice_pPb_double_ridge, pPb_mass_ordering, multiparticleCorr_pPb, p_pi_enhancement_pPb, CMS_pp_Ridge, d-Au ridge paper}. Even
hydro-based models  \cite {epos_massordering_flow_pPb}, \cite {epos_ridgein_pp}, \cite {epos_radialflow_spectra_pPb} found to be  in reasonable agreement with experimental results indicating that local thermal equilibration might be achieved  even for small system size. In \cite {AMPT_pp_pPb_ridge}, \cite {AMPT_pPb_flow} it is shown that microscopic transport models(AMPT) are also capable of generating similar effects in small collision systems  through in-coherent parton scattering with a nominal scattering cross-section.
The system evolving in transport model is relatively less dense [12] compared to the system evolving in near hydro limit where large number of collisions among the constituents generate the pressure gradient and the hydro like collectivity. In \cite {AMPT_escape_mech} it has been shown that anisotropic escape mechanism of partons is the dominant source of flow ($v_{2}$) generation in AMPT. 
Recent studies \cite {AMPT_mass_splitting} also indicate that mass-splitting of $v_{2}^{hadron}$ in AMPT
originates from the dynamics of coalescence \cite {Lin_Coalescence_dynamics} and hadronic re-scattering during the evolution of the system and not necessarily associated with the collectivity in the system.
\\
\\
In this article we aim to present further-insight on the possible role of partonic and hadronic dynamics in the generation of mass ordering of $v_{2}^{hadron}$ in AMPT.
Events generated from the default and string-melting version of AMPT for different scattering cross-sections have been analysed to calculate elliptic flow of partons and hadrons (pions and (anti-)protons) as a function of event-activity (multiplicity).
Although anisotropic flow in AMPT is dominantly from the escape mechanism but its qualitative features exhibit striking
similarity with collective response of the medium.

Our study based on the AMPT (SM version) generated data for p-Pb collisions at $\sqrt{s_{NN}}$ = 5.02 TeV suggests that the parton cascade coupled with dynamics of coalescence \cite {Lin_Coalescence_dynamics}, \cite {RFPRL90_2003_14}, \cite {VGPRL90_2003_15}  can generate mass ordering of $v_{2}^{hadron}$  at the partonic level without hadronic interactions. Also, hadronic interactions alone can generate mass ordering of $v_{2}^{hadron}$ without any contribution from the partonic phase and therefore the total effect has a contribution from both partonic and hadronic level. 

%Our study suggests that the initial phase space distribution of the partons need to be modified before %hadronization by coalescence \cite {Lin_Coalescence_dynamics}, \cite {RFPRL90_2003_14}, \cite %{VGPRL90_2003_15} to generate the mass ordering of $v_{2}^{hadron}$ in AMPT. 

% Interestingly, model calculation also reproduces mass-ordering in the pT-differential elliptic anisotropy of %the final state particle. Mass splitting in the elliptic flow measurements is natural to hydrodynamics but %kinematic models with modest scattering cross-section are far-away from the hydro-limit. This %observation lead to an important conclusion that mass ordering is not  uniquely associated with %collective flow of hydrodynamical origin.
%In small systems where hadronic interactions may be ignored due to low hadron density, mass ordering %at low pT can be solely generated through quark-coalescence.

\section{The AMPT Model}

The AMPT model~\cite{ampt,ampt2,ampt3} has been extensively used to study collision dynamics at relativistic energies.
It is a hybrid transport model that comprises of four
major components: the fluctuating initial conditions, a cascade model for partonic
interactions, transition from partonic to hadronic degrees of freedom and
final state hadronic evolution. 
The  model provides two modes of operations: Default and String Melting (SM). 
In both cases, the initial distribution of mini-jet partons and strings are derived from the HIJING~\cite{ampt4}.
In the default version, only minijet partons are subjected to partonic scattering modeled by Zhang Parton Cascade (ZPC)~\cite{ampt5} model.
Currently this model includes only 2-body elastic scatterings with scattering cross-section derived from the leading order p-QCD
calculations as:
\begin{eqnarray}
 \sigma = \frac{9\pi \alpha^2_s}{2\mu^2}
\end{eqnarray}
In Eq.1, $\alpha_s$ is the strong coupling constant and $\mu$ 
is the Debye screening mass. In this analysis, two scattering cross sections - 0 mb (ZPC off) and 3 mb (ZPC on) have been used. At the completion of the cascade, minijet partons are recombined with their parent strings and hadronized
via the Lund string fragmentation, using the following fragmentation function~\cite{ampt6, ampt7}:
\begin{eqnarray}
f(z) \propto~ z^{-1}(1-z)^{a}exp(-b m^2_{T}/z),
\end{eqnarray}
where the  fragmentation parameters $a$ and $b$ are 
set to 0.5 and 0.9 GeV$^{-1}$ for this analysis.

In string melting version, excited strings from the HIJING are converted to valance quarks
and anti-quarks which are fed to ZPC for space-time evolution.
The quarks and anti-quarks at end of ZPC are converted to hadrons using a spatial coalescence
model. Both in default and string melting version, subsequent hadronic evolution is modeled by
A Relativistic Transport model~\cite{ampt8}.
In this work , we have determined the momentum space azimuthal anisotropy of the freezeout partons after ZPC just before hadronization. After hadronization the $v_{2}$ of hadrons is determined with and without hadronic rescatterings. 

\begin{figure}[htb!]
\begin{center}
\includegraphics[width=0.40\textwidth]{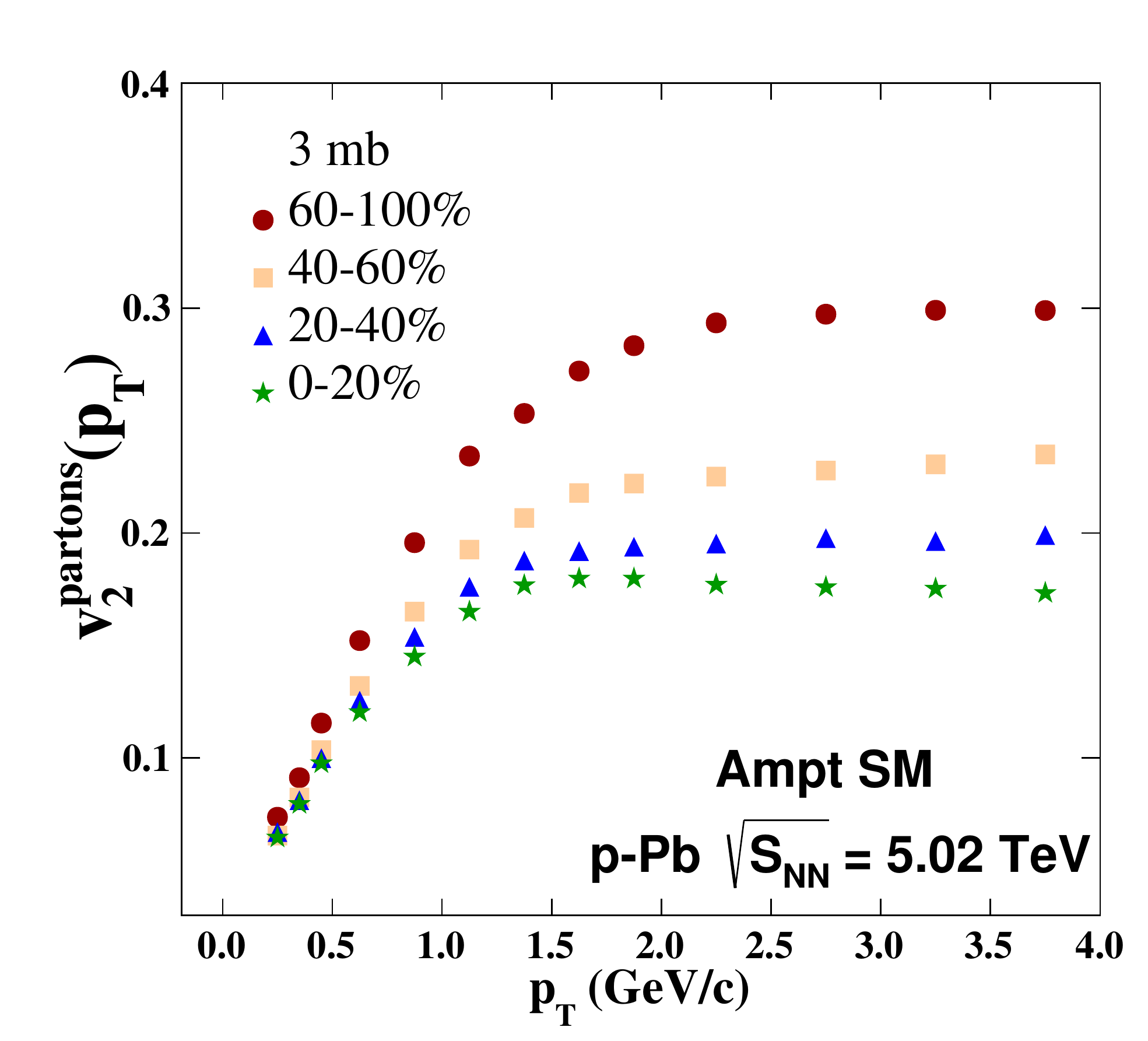}
\caption{[Color online] Multiplicity dependence of $p_{T}$-differential $v_{2}^{parton}$ in p-Pb collisions at $\sqrt{s_{NN}}$ = 5.02 TeV from AMPT-SM
with parton scattering cross-section of 3 mb.}
\end{center}
\label{v2parton_diffCross}
\end{figure}

\begin{figure}[htb!]
\begin{center}
\includegraphics[scale=0.32]{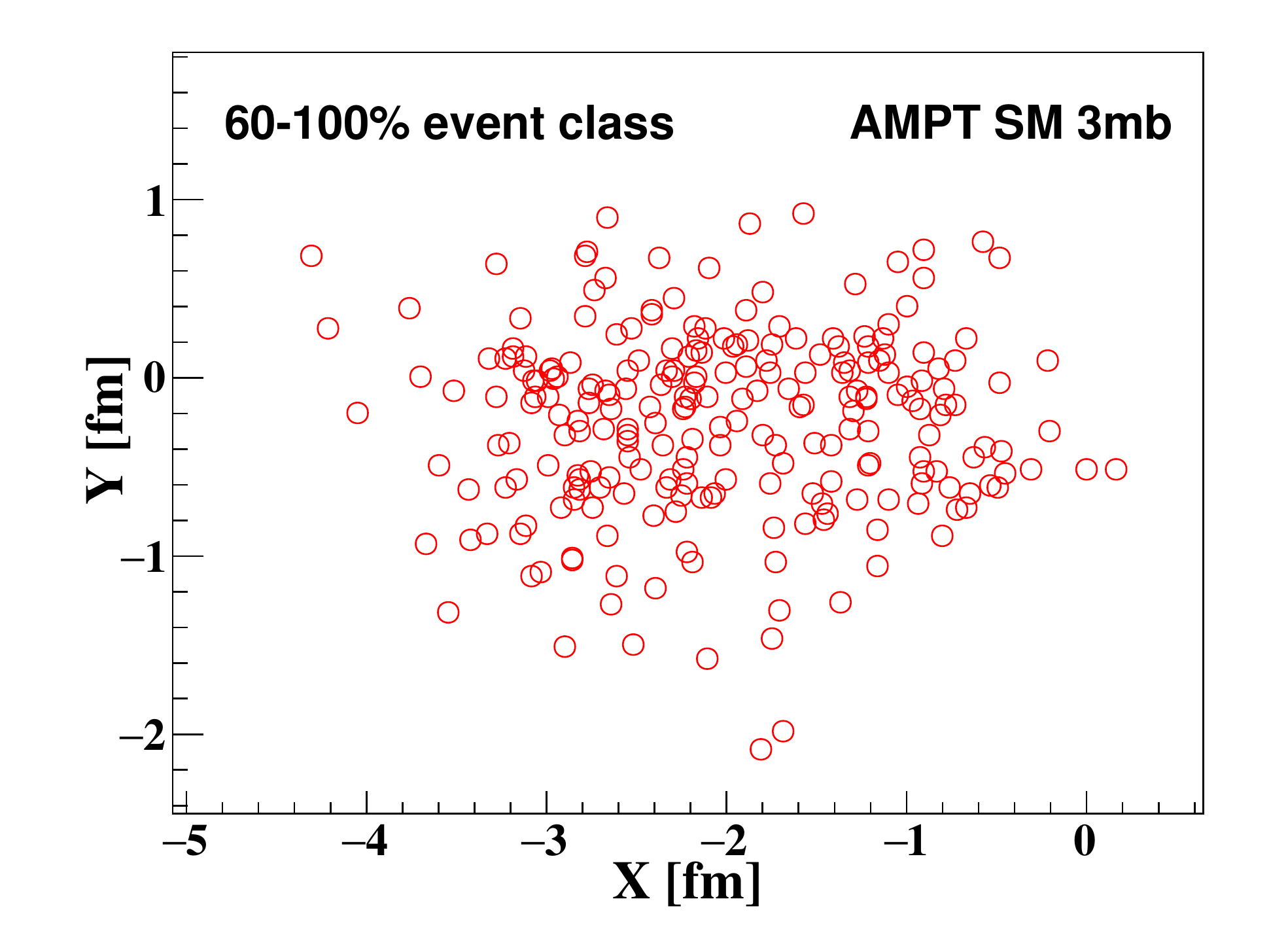}
\includegraphics[scale=0.32]{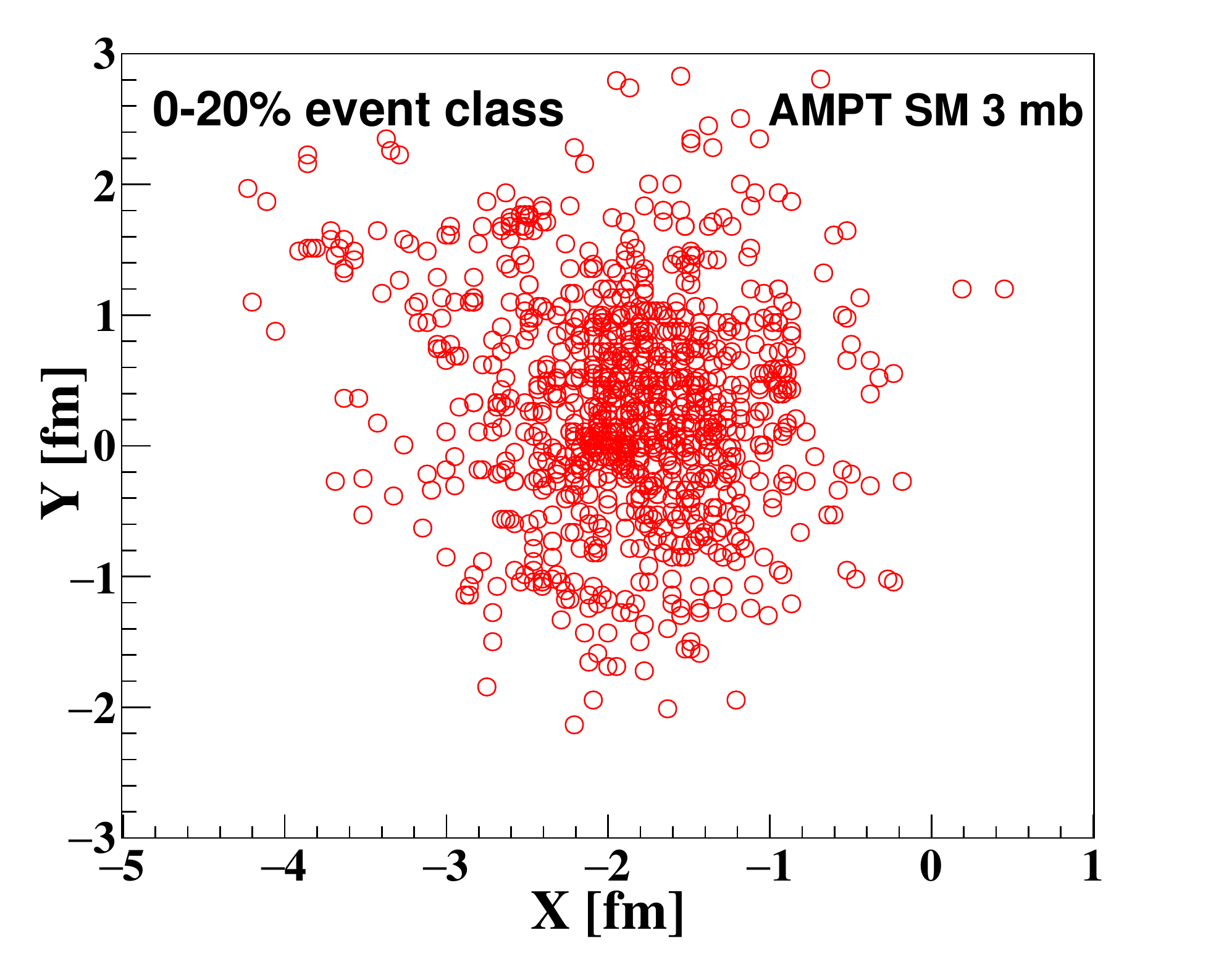}
\caption{[Color online] Coordinate space distribution of freezeout partons in a single event in
 a) 60-100\% (top) and  b) 0-20\% (bottom) event class of p-Pb collisions at $\sqrt{s_{NN}}$ = 5.02 TeV.}
\end{center}
\label{partondist_XY}
\end{figure}

\section{$\bf{v_{2}}$ extraction}
The anisotropic emission of charged particles can be quantitively charecterised in terms of the co-efficients in the Fourier
expansion of the azimuthal dependence in the invarient yield  relative to the reaction plane angle~\cite{Sec2_ref1}:
\begin{equation}
 E\frac{d^{3}N}{dp^{3}} =\frac{1}{2\pi}\frac{d^{2}N}{p_{T}dp_{T}dy}(1 +\sum_{1}^{n} 2v_{n}cos(n(\phi - \psi_{R}))
\end{equation}
 Where $\phi$ is the azimuthal angle in the Lab-frame and $\psi_{R}$ is the reaction plane angle.
The second term in the expansion, $v_{2}$, is known as elliptic flow and can be simply obtained by $<<cos(2(\phi - \psi_{R})>>$.
The angular brackets stands for the statistical average over many events.
Under the experimental conditions reaction plane $\psi_{R}$ can  not  be determined directly hence new methods have been  

\begin{figure}[htb!]
\begin{center}
\includegraphics[scale=0.36]{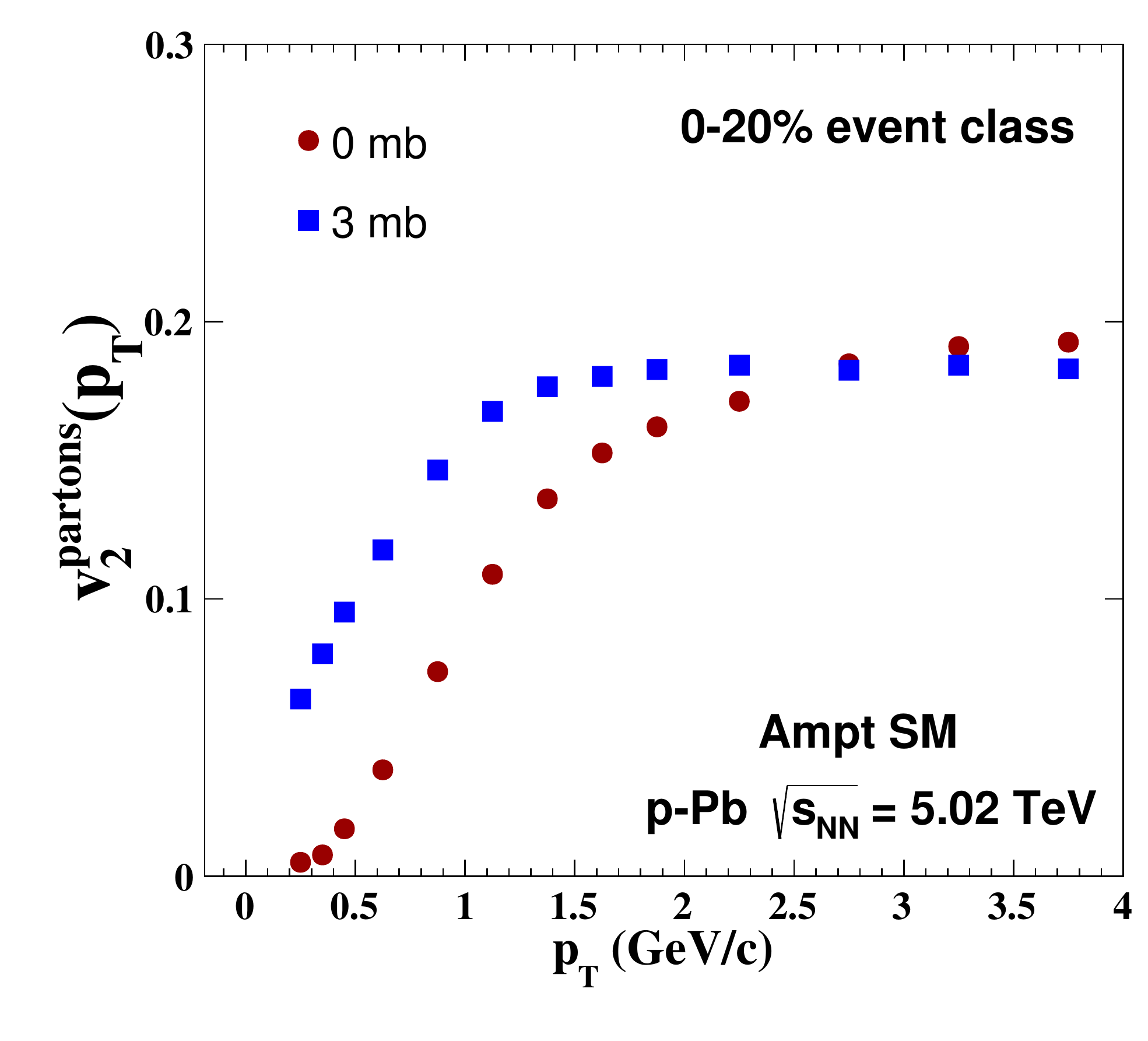}

\caption{[Color online] $v_{2}^{parton}$ plotted as a function
of $p_{T}$  in the highest multiplicity event class (0-20\%) of p-Pb collisions at $\sqrt{s_{NN}}$ = 5.02 TeV for parton scattering cross-section of 0 mb (red) and 3 mb (blue) from AMPT-SM version.}
\end{center}
\label{v2_parton}
\end{figure}

improvised to calculate $v_{2}$
independent of the reaction plane angle. The azimuthal correlations among two or multi particles 
~\cite{Sec2_ref2,Sec2_ref3,Sec2_ref4} is found to be an useful tool
 to reconstruct $v_{n}$ coefficients. Thus one can calculate $v_{2}$ from two-particle azimuthal correlations as
~\cite{Sec2_ref5}:
\begin{equation}
 <<e^{i2(\phi_{i}-\phi_{j})}>> = <v_{2}^{2}> + \delta_{n}
\end{equation}
$\delta_{n}$ in the above equation is the ``non-flow'' contribution to the two particle correlation. In the high multiplicity events this calulation may be computationally intensive due to large number of 2-particle combinations.
To counter computational in-efficiency a revision to this method was suggested called as Q-cumulant method ~\cite{Sec2_ref5,Sec2_ref7}
. In this approach, 2-particle
correlations are expressed in terms of flow vectors or Q-vectors mathematically represented as:
\begin{equation}
 Q_{2} = \sum_{1}^{M} e^{i2\phi_{i}}
\end{equation}
Summation runs over all particles usually called as reference particles (RP). The genuine 2-particle azimuthal correlations can be obtained by
separting the diagonal and the off-diagonal terms in $|Q_{2}|^{2}$ as:
\begin{equation}
 |Q_{2}|^{2}= M + \sum^{'} e^{i2(\phi_{i} - \phi_{j})}
\end{equation}
Thus the average 2-particle azimuthal correlations over all particles (RP) in a single event can be calculated using:
\begin{equation}
 <2> = \frac{|Q_{2}|^{2} - M} {M (M -1)}
\end{equation}
$<>$ represents particle average in a single event. Finally the 2-particle cumulant $c_{2}\{2\}$ and $\mathrm{V_{2}}\{2\}$ can be obtained averaging over all particles over
all events:
\begin{equation}
 c_{2}\{2\} = <<2>> ,
\mathrm{V_{2}}\{2\} = \sqrt{<<2>>}
\end{equation}
$<< >>$ indicates both particle and event average.\\
$v_{2}$ obtained in this method is prone to contaminations from non-flow effects like decay of resonances,
jet induced correlations etc. To suppress these additional correlations which are generally short range, an optimum 
$\Delta\eta$ gap can be introduced between reference particles (RP)~\cite{Sec2_ref8}. This can be achieved by divding an event into two sub events, A and B , separated by $\Delta\eta$ gap.
Hence the Eq (7) is modified as :
\begin{equation}
 < 2 >_{\Delta\eta} = \frac{Q^{A}_{2}.Q^{B*}_{2}}{M_{A}.M_{B}}
\end{equation}
Where $ Q^{A}_{2}, Q^{B}_{2} $ are flow vectors corressponding to the sub-events A and B , and $M_{A} , M_{B}$ are the multiplicities of 
the RP's in each sub-events.
Having calucated integrated flow ($\mathrm{V_{2}}\{2\}$) one can extract differential flow of Particles of Interest (POI's)
using the analogus technique. In sub-event Q-cumulant method, differential flow of particles can be expressed as:
\begin{equation}
 <2^{'}_{\Delta\eta}> = \frac{p_{2,A}. Q_{2}^{B*}}{m_{p,A}. M_{B}}
\end{equation}
Where $m_{p,A}$ is the number and $p_{2}$ is the Q-vector of the particles of interest (POI's) whose differntial flow calculation is intended . To avoid overlap in the
psudorapidity range among POI's and RP's they are taken from different sub-events.
Finally, differntial flow is calculated via:
\begin{equation}
 v_{2}\{\it{{p_{T}}}\} = \rm\frac{<2^{'}_{\Delta\eta}>}{V_{2}\{2\}}
\end{equation}
In this work, all freezeout partons / final state hadrons produced within the pseudorapidity range of $|\eta| < 1.0$ are   considered. Each event is divided into two
sub-events with a  pseudo-rapidity separation among the freezeout partons / final state hadrons of ($|\Delta\eta|$) $>$ 0.4 at least. 

This analysis (extraction of  $v_{2}^{parton}$ and  $v_{2}^{hadron}$) has been performed by dividing the entire minimum bias events
into four multiplicity classes based on the total amount of charged particles produced (with $p_{T}$ $>$0.05 GeV/c) within 2.8  $<\eta<$5.1. This is the acceptance range of ALICE VZERO-A detector in the Pb going direction in case of p-Pb collisions and used for multiplicity class determination by the ALICE collaboration \cite {alice_pPb_double_ridge},\cite {pPb_mass_ordering}. The multiplicity classes are denoted as 60-100$\%$, 40-60$\%$, 20-40$\%$, 0-20$\%$ from the lowest to the highest multiplicicty.

\section{Results and Discussion}

In this paper the multiplicity evolution of the elliptic flow for freezeout partons ($v_{2}^{parton}$) and hadrons ($v_{2}^{hadron}$) in p-Pb collisions  
at $\sqrt{s_{NN}} = $ 5.02 TeV has been studied. The effect of partonic interactions via Zhang's parton cascade (ZPC) and different hadronization mechanisms (Lund string fragmentation (LSF) in default and coalescence  in string melting (SM) version of AMPT)
on mass splitting of $v_{2}^{hadron}$ has been investgated. The effect of hadronic interactions on the mass splitiing of $v_{2}^{hadron}$ has also been reported.

In Fig 1 the multiplicity evolution of the elliptic flow of freezeout partons ($v_{2}^{parton}$)  is shown with ZPC on (scattering cross section of 3 mb). The $v_{2}^{parton}$ decreases with increase in multiplicity. 
In AMPT the freezeout partons exhibit space momentum correlations largely due to the escape mechanism \cite {AMPT_escape_mech} and the escape probability of a freezeout parton depends on it's position and momentum at the freezeout point. The escape probability of these freezeout partons may be affected 
by the phase space distribution of the surrounding partons that depends on the intial geometry of the system - influencing the preferential escape direction of the partons along the shorter axis. In the lowest multiplicity class (60-100\%) of p-Pb collisions where the partonic interactions are less significant, the non flow effects \cite {zhou_urqmd} and escape mechanism \cite {AMPT_escape_mech} generate larger $v_{2}^{parton}$ compared to the higher multiplicity classes where $v_{2}^{parton}$ has additional contribution from the space-momentum correlation generated during partonic interactions during ZPC.

In Fig 2. the coordinate space distributions of the freezeout partons in a single event are shown for the highest(0-20$\%$) and 
lowest (60-100$\%$) multiplicity classes. The large density of partons in the highest multiplicity class possibly 
reduces the non flow effects and escape probability of the freezeout partons, thus reducing the $v_{2}^{parton}$ compared to the lowest multiplicity class as shown in Fig 1. It indicates that even though the collective contribution to the $v_{2}^{parton}$ increases in the higher multiplicity classes due to larger number of collisions among partons, the total $v_{2}$ goes down as the escape probability of the partons and non flow effects decrease with increase in multiplicity.
%The scenario is qualitatively same in the ZPC on case as the multiplicity evolution of $v_{2}^{parton}$ is %similar in both cases.
In Fig 3  the $p_{T}$ dependence of $v_{2}^{parton}$  for ZPC on (scattering cross section of 3 mb) and off (scattering cross section of 0 mb) in SM version of AMPT is shown for the highest multiplicity event class. In ZPC off case, the $v_{2}^{parton}$ is solely due to non flow effects \cite {zhou_urqmd} as there is 
no partonic interactions present. Whereas, in ZPC on case the escape mechanism and hydro like collectivity generated through partonic interactions contribute towards the observed $v_{2}^{parton}$. In both cases the $v_{2}^{parton}$ is approximately linear at low $p_{T}$ and it saturates at higher $p_{T}$.

In this paper we have investigated the effect of different
hadronization mechanisms (with and without ZPC) on the mass splitting of $v_{2}^{hadron}$ which was believed to be a characteristic
signature of hydro like collectivity. A recent study has shown that the mass ordering can be generated from the dynamics of the coalescence \cite {Lin_Coalescence_dynamics} in AMPT even when parton azimuthal directions are randomized \cite {AMPT_mass_splitting}. In Fig 4 the multiplicity evolution of the mass splitting of $v_{2}^{hadron}$ in string melting(SM) version of AMPT for ZPC on (with and without hadronic scattering) is shown. In case of ZPC on , the mass splitting is evident even in the lowest multiplicity event class and this splitting increases with increase in multiplicity whereas, the value of  $v_{2}^{hadron}$ decreases. As the number of collisions

\begin{figure}[htb!]
\begin{center}
\includegraphics[width=0.45\textwidth]{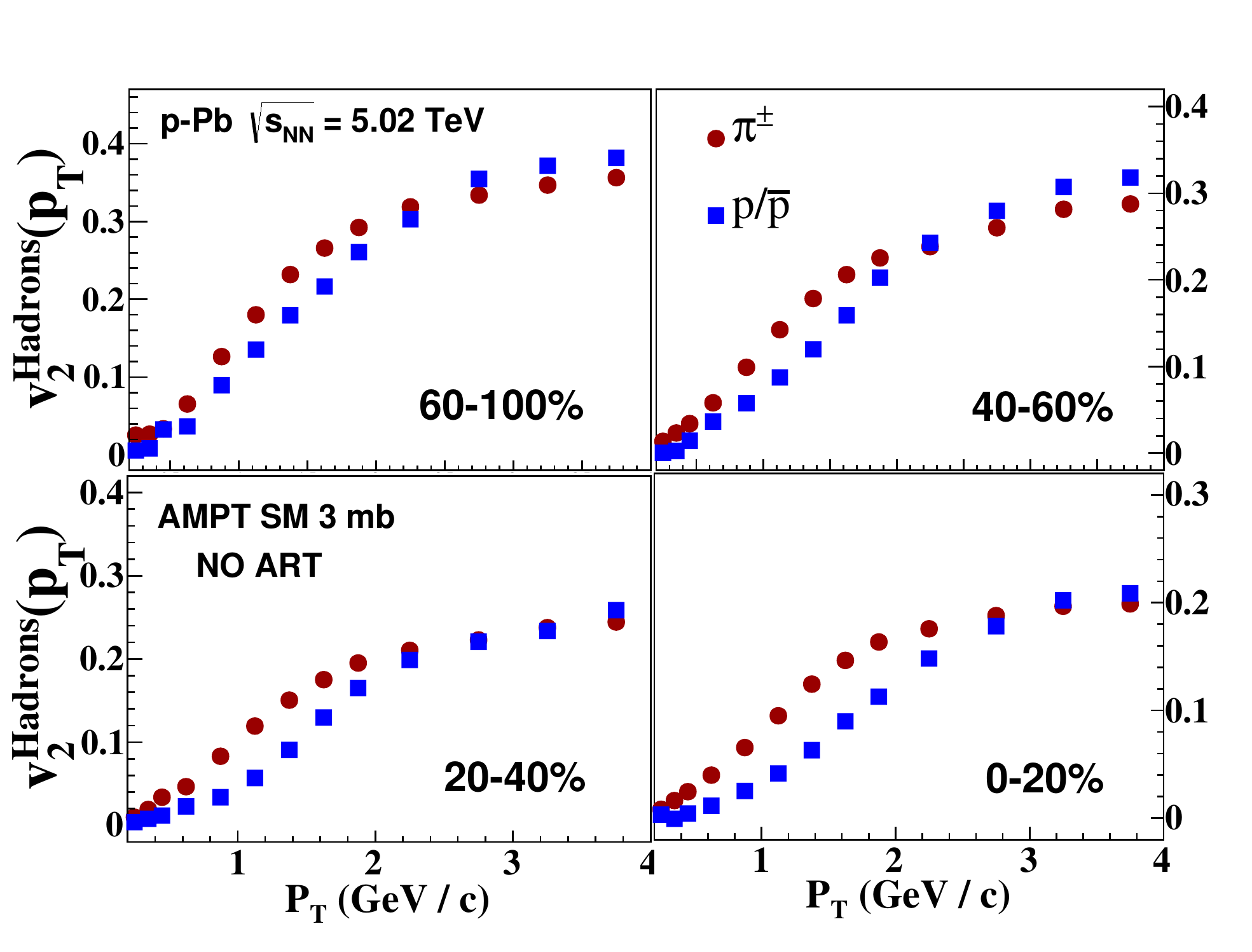}
\includegraphics[width=0.45\textwidth]{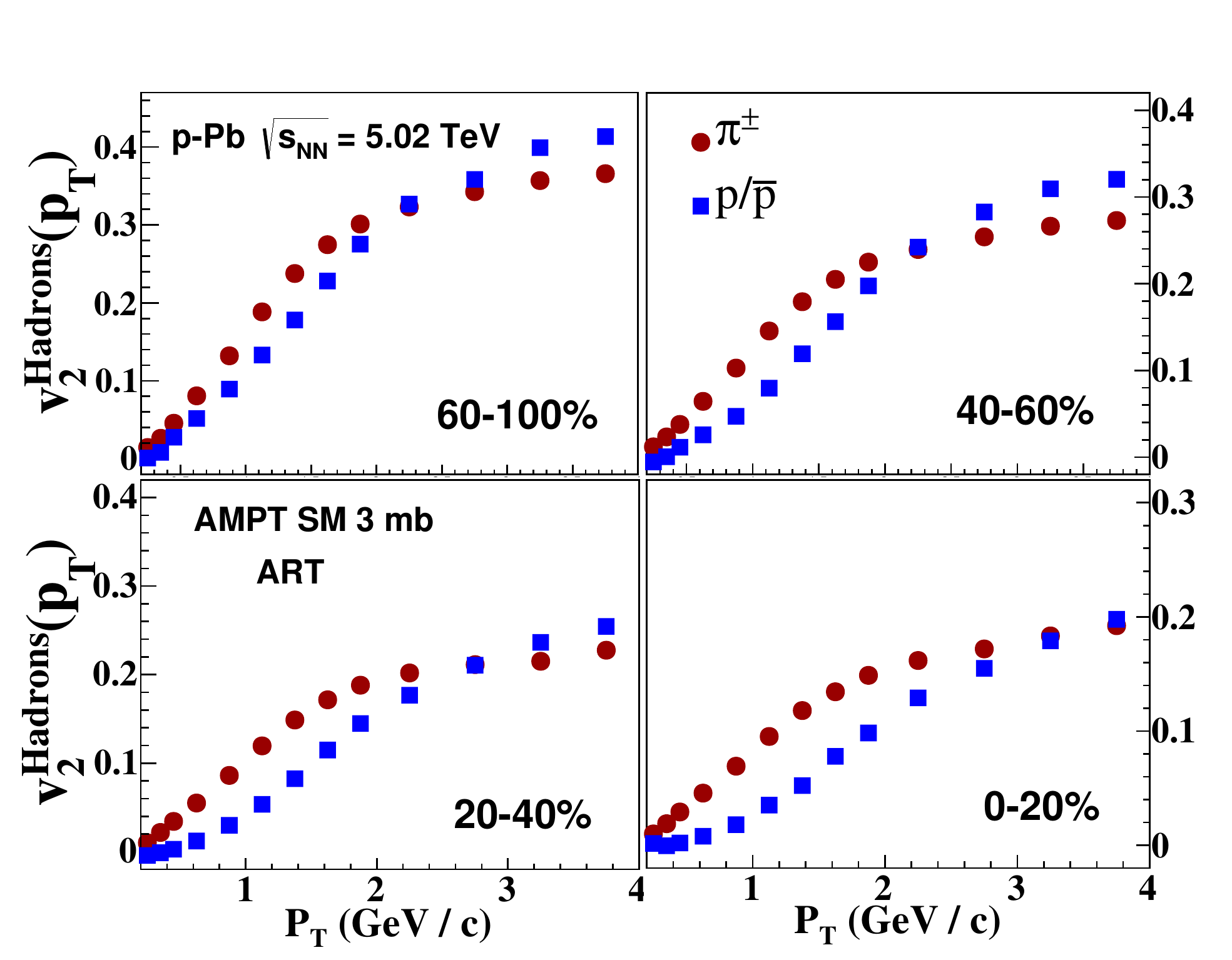}

\caption{[Color online] Multiplicity evolution of  $v_{2}(p_{T})$ of pion and proton for 3 mb parton
scattering cross-section in p-Pb collisions at $\sqrt{s}$ = 5.02 TeV with (bottom plot)
and without (upper plot) hadronic-rescattering (ART).}
\end{center}

\end{figure}

\begin{figure*}[htb!]
%\begin{center}
\includegraphics[scale=0.26]{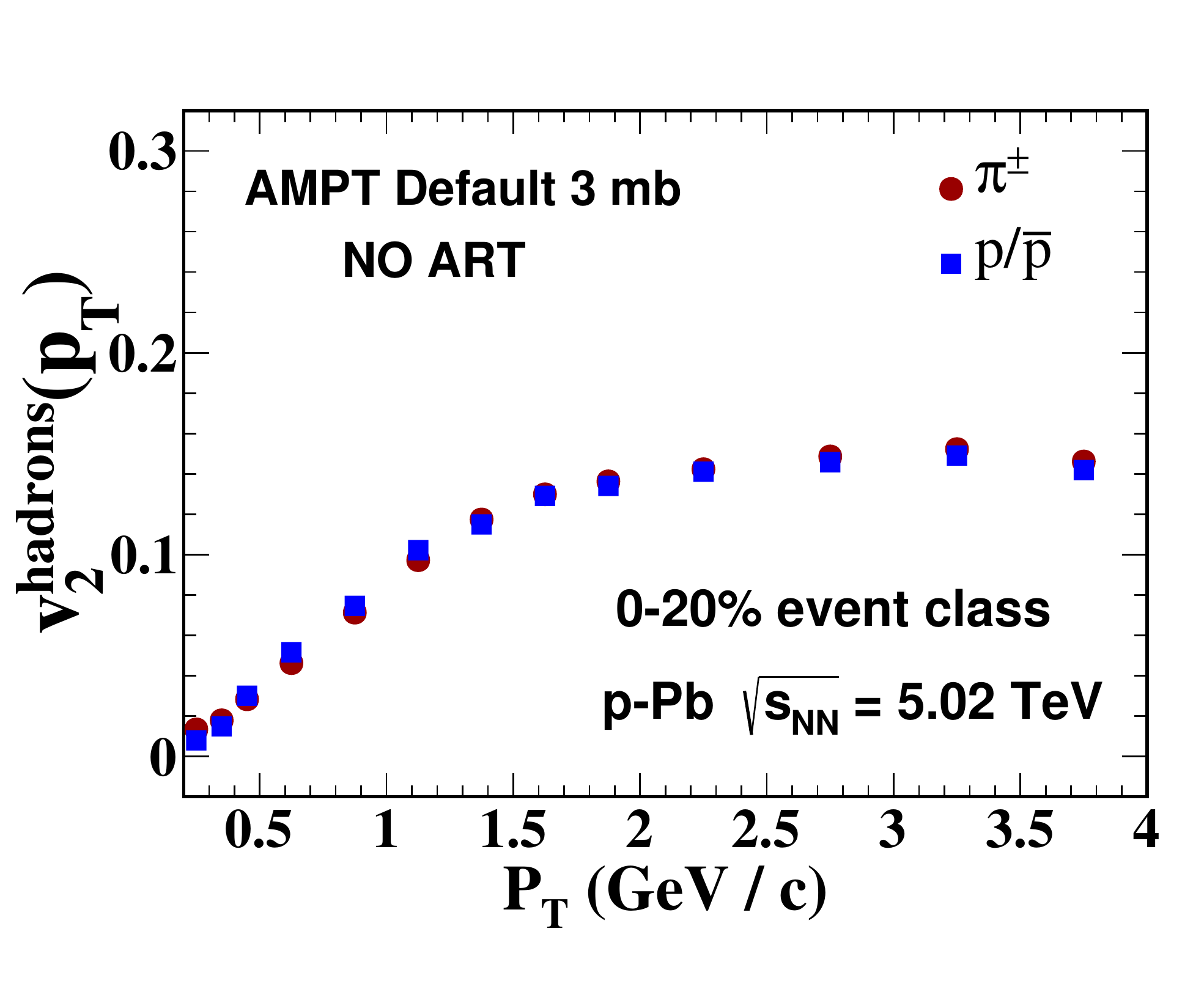}
\includegraphics[scale=0.26]{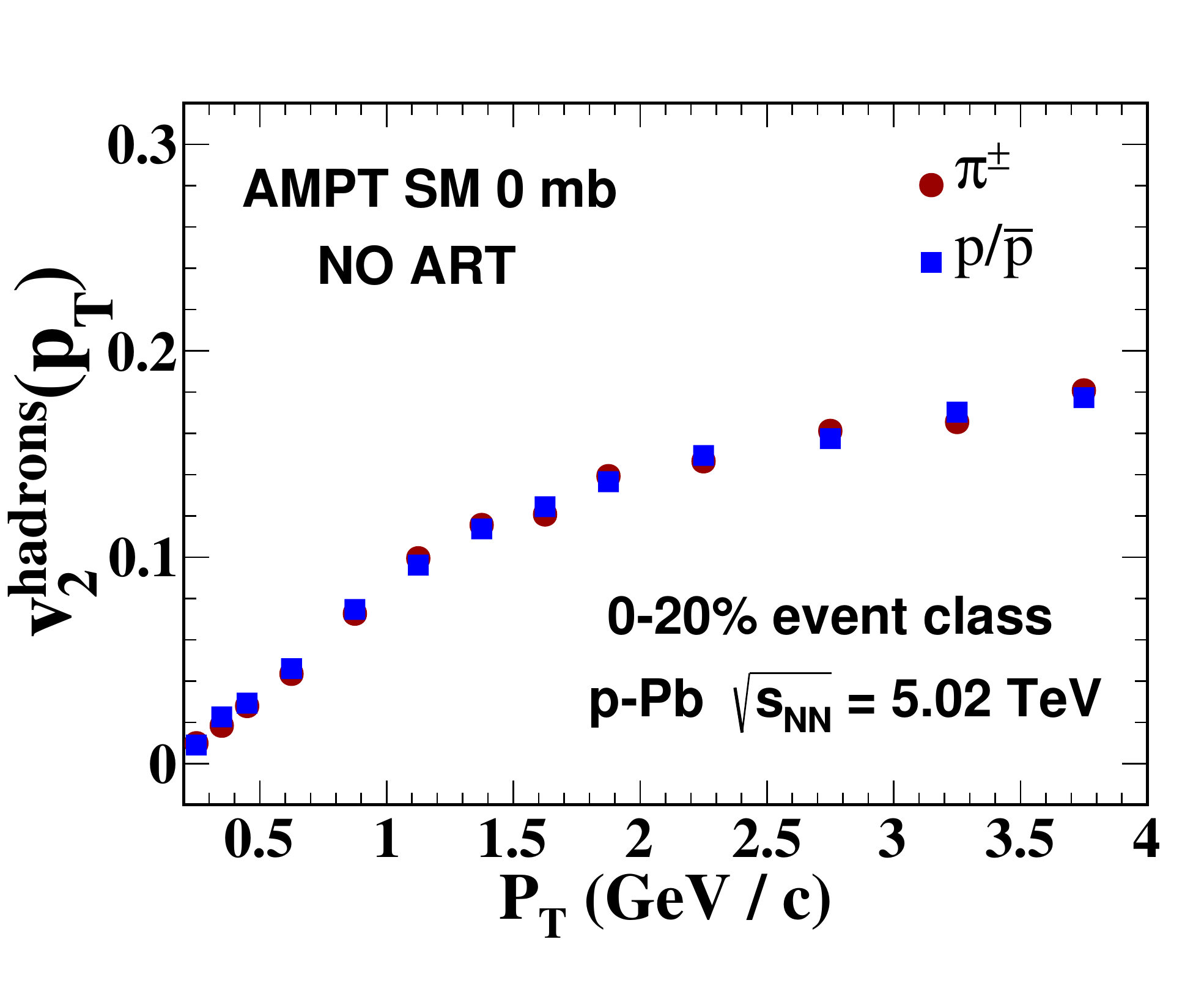}
\includegraphics[scale=0.25]{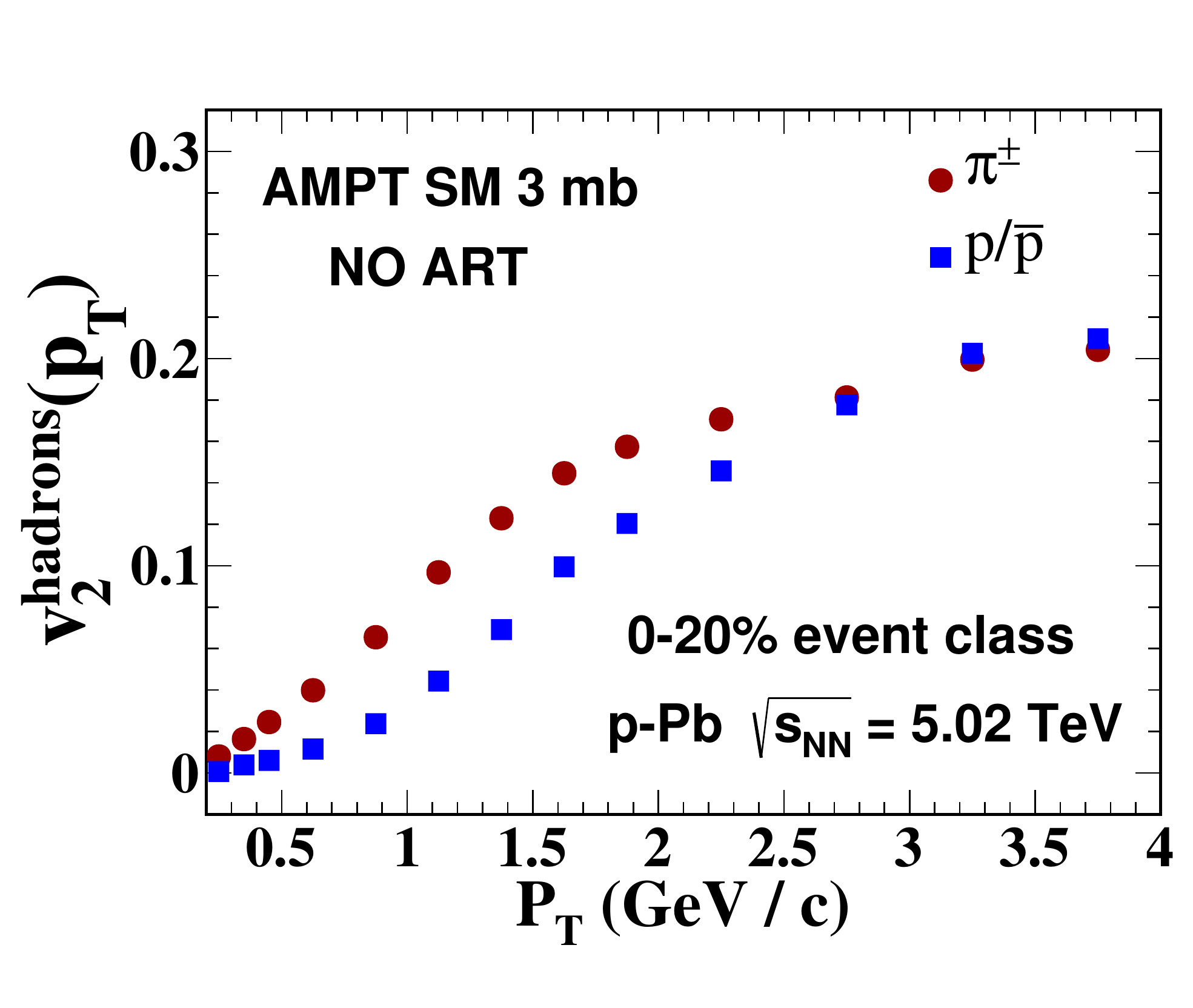}
\caption{[Color online] $v_{2}(p_{T})$ of pion and proton for most central (0-20\%)
event class in p-Pb collision at  $\sqrt{s_{NN}}$ = 5.02 TeV for different configurations: a) Default 3 mb without ART (left),
b) 0 mb String Melting without ART (center) and c) 3 mb String Melting without ART (right).}
%\end{center}

\end{figure*}

\begin{figure*}[htb!]
%\begin{center}
\includegraphics[scale=0.26]{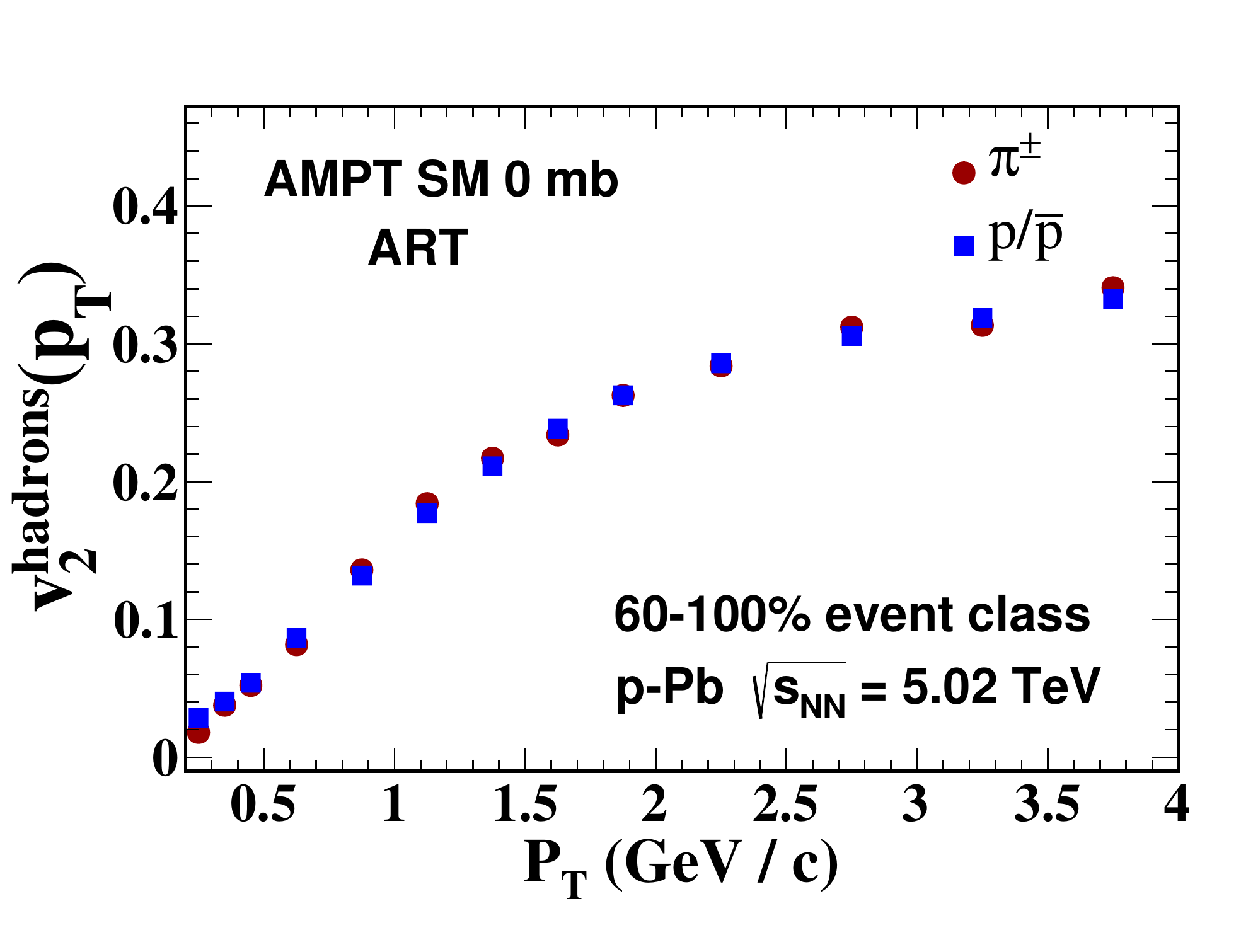}
\includegraphics[scale=0.27]{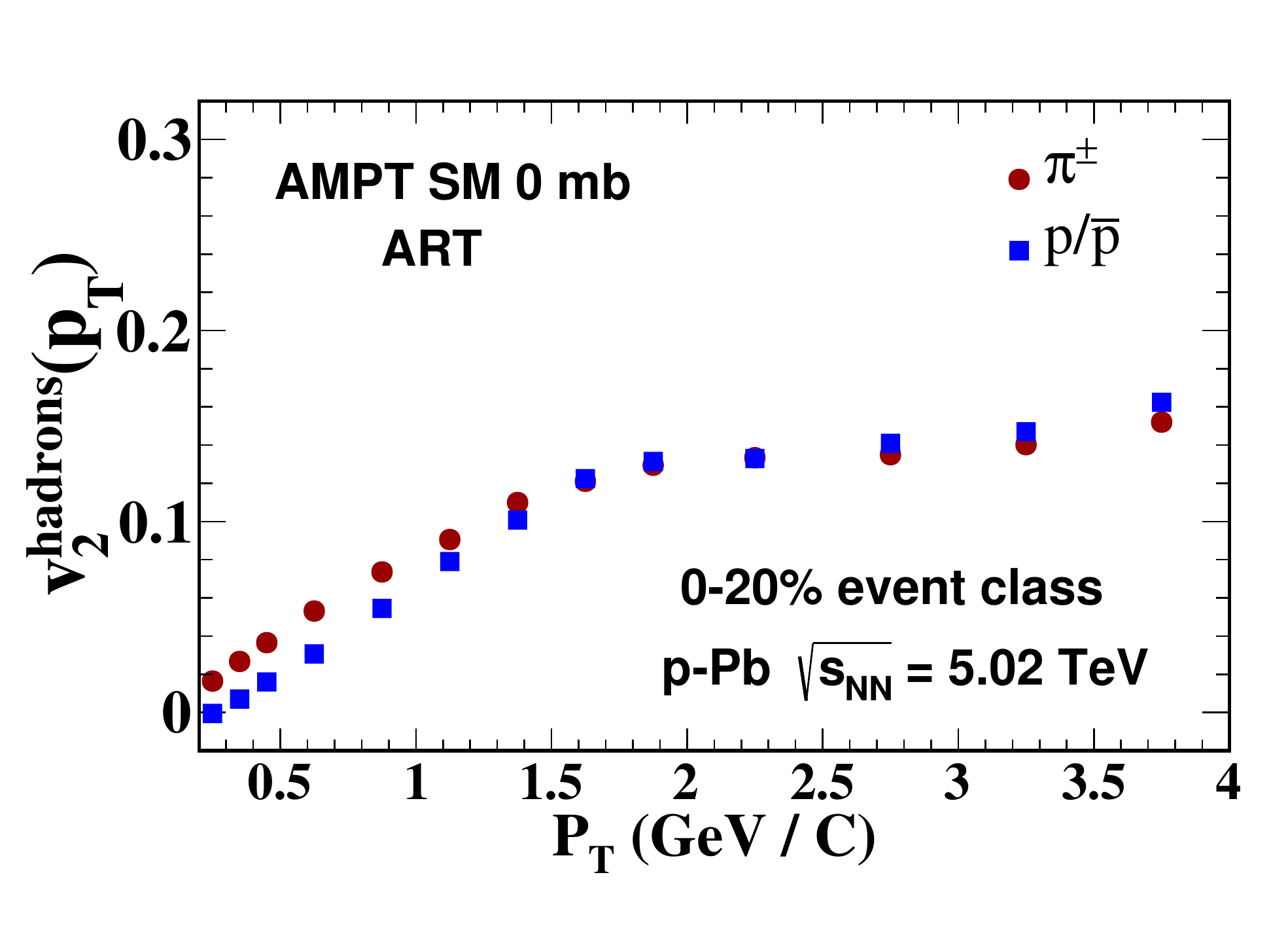}

\caption{[Color online] $v_{2}(p_{T})$ of pion and proton in a) 60-100\% (left) and  b) 0-20\% (right) event class of p-Pb collisions at $\sqrt{s_{NN}}$ = 5.02 TeV in string melting version of AMPT with ART on and ZPC off.}
%\end{center}

\end{figure*}
suffered by a parton increases with multiplicity, the collective contribution to the elliptic anisotropy  increases resulting in more prominent mass splitting. However if ZPC is turned off, the coalescence model of hadronization  is unable to produce any mass splitting of $v_{2}^{hadron}$ even in the highest multiplicity event class as shown in Fig. 5(b) ({\it AMPT SM 0 mb}). In ZPC off case the $v_{2}^{parton}$ is solely due to non flow effects \cite {zhou_urqmd} and it's $p_{T}$ dependence is similar to the ZPC on case as shown in Fig. 3. Due to  linearity of $v_{2}^{parton}$ at low $p_{T}$, if constituent quarks $p_{T}$ and $v_{2}$ simply add to hadronic $p_{T}$ and $v_{2}$ then baryon and meson $v_{2}$ should coincide with each other and there would be no mass splitting at low $p_{T}$ \cite {AMPT_mass_splitting}. Now, even though the linearity of  $v_{2}^{parton}$ at low $p_{T}$ is present in both ZPC on and off cases - after coalescence it is only the ZPC on case where mass splitting of $v_{2}^{hadron}$ is observed even in the lowest multiplicity event class as
shown in Fig. 4.  Whereas, no mass splitting is observed in ZPC off case (Fig 5 (b)) even in the highest multiplicity class. The parton cascade changes the initial phase space distribution of the partons. So during coalescence after ZPC, the constituent parton $p_{T}$  may have a spread in momentum for a given hadron $p_{T}$ and $v_{2}^{parton}$ do not add up arithmatically to hadron $v_{2}$ because of finite opening angles or kinematics \cite {AMPT_mass_splitting}. The dynamics of coalescence  \cite {Lin_Coalescence_dynamics} among these constituent partons (having a spread in their momentum and finite opening angles) is considered to be the source of mass splitting of $v_{2}^{hadron}$ at low $p_{T}$ in AMPT \cite {AMPT_mass_splitting} before any hadronic scatterings take place. 
Now, parton cascade creates space-momentum correlation. Whereas, in \cite {AMPT_mass_splitting} it is argued that the dynamics of coalescence can generate the mass ordering in the azimuth-randomized version of AMPT where the parton space-momentum correlation is destroyed by randomizing parton azimuthal directions. In that case also 
 %This is also in agreement with the observation made in \cite {AMPT_mass_splitting}. In that case also
the initial phase space distribution of the partons are modified by ZPC followed by ranomizing the outgoing parton azimuthal directions after each parton-parton scattering \cite {AMPT_escape_mech} \cite {AMPT_mass_splitting}, and then the dynamics of coalescence creates the mass ordering of $v_{2}^{hadron}$ even in the absence of any event plane correlation.\\
To study further, we repeated our analysis with default version of AMPT keeping ZPC on without hadronic scattering. In Fig 5, the comparison between 3 configurations - {\it default 3 mb} (ZPC + Lund string fragmentation (LSF)) , {\it SM 0 mb} (No ZPC + coalescence (SM)) and {\it SM 3 mb} (ZPC + coalescence (SM)) 
has been shown for the highest multiplicity event class. No mass splitting is observed for both 
{\it default 3 mb} (Fig 5(a)) and {\it SM 0 mb} Fig(5(b)) case. This suggests that the parton cascade (ZPC) combined with lund string fragmentation and coalescence without parton cascade are unable to produce any mass splitting of $v_{2}^{hadron}$ without hadronic scattering- confirming that both ZPC and coalescence model of hadronization at the partonic level (Fig 4 and Fig 5(c)) are one of the sources of the observed mass splitting of $v_{2}^{hadron}$.

The mass splitting is found to be slightly enhanced under the influence of hadronic interactions mainly in the highest multiplicity event class as shown in Fig.4. To study the effect of hadronic scattering alone, we repeat the analysis with hadronic scattering (ART) only with partonic scattering turned off in SM version of AMPT. In the lowest multiplicity (60-100\%) class, no mass splitting of $v_{2}^{hadron}$ is observed as shown in Fig 6(a). But, in the highest multiplicity class (0-20\%), as shown in Fig 6(b), a clear mass splitting is observed at lower $p_{T}$. This indicates that with the increase in hadronic density, contribution from hadronic interactions towards mass splitting increases significantly, which may play a major role in case of heavy ions. This measurement is in agreement with the observations made in  \cite {AMPT_mass_splitting} and \cite {zhou_urqmd}.\\ 
Our study based on the AMPT model suggest that in small collision system like p-Pb, dynamics of coalescence can generate the mass splitting of $v_{2}^{hadron}$ even in the lowest multiplicity class of p-Pb collisions in absence of any hadronic scattering, provided partonic interactions are allowed prior to hadronization via coalescence mechanism.  Also, in Fig 6 it is shown  that hadronic scattering alone can generate the mass splitting of $v_{2}^{hadron}$ in the high multiplicity class of p-Pb collisions consistent with the observations reported in \cite {AMPT_mass_splitting} and \cite {zhou_urqmd}. Our observations indicate that the mass ordering of $v_{2}^{hadron}$ in AMPT can originate independently from both partonic (ZPC+coalescence) and hadronic (ART) phase. 
The partonic contribution can generate this effect even in the lowest multiplicity class of p-Pb collisions (Fig 4) where the system is expected to be far away from the ideal hydro limit \cite {AMPT_mass_splitting}. Whereas, interactions at both partonic and hadronic phase can generate the mass ordering in the higher multiplicity classes (Figs 4 and 6). This suggest that the mass splitting of $v_{2}^{hadron}$ is not uniquely associated with the hydrodynamical evolution of the partonic phase (QGP), produced in relativistic high energy collisions.

%In the AMPT model in use, if partonic scattering (ZPC) is on, coalescence can generate the mass %ordering even in the lowest multiplicity event class of p-Pb without any hadronic scattering as shown in %Fig. 4. Also, in fig 5(b) it is shown that without ZPC coalescence can not generate the mass splitting of %$v_{2}^{hadron}$.  Now, parton cascade creates the event plane correlation. whereas, in \cite {%AMPT_mass_splitting} it is argued that the dynamics of coalescence can generate the mass ordering in %the azimuth-randomized simulations. In that case also 
 %This is also in agreement with the observation made in \cite {AMPT_mass_splitting}. In that case also
%the initial phase space distributions of the partons are modified by ZPC followed by ranomizing the %outgoing parton azimuthal directions after each parton-parton scattering \cite {AMPT_escape_mech} \cite %{AMPT_mass_splitting} and then the dynamics of coalescence creates the mass ordering of %$v_{2}^{hadron}$ even in the absence of any event plane correlation.

\section*{Acknowledgements} 
 Thanks to VECC grid computing team for their 
constant effort to keep the facility running and helping in AMPT data generation.

%\linenumbers
\end{document}